\documentclass[twocolumn,trackchanges]{aastex701}

\usepackage{siunitx}

\begin{document}

\title{Apparent Transverse Motion of Light Bridges Coupled to Coronal Loop Dynamics}

\author[orcid=0000-0003-1619-7410,gname=Atul,sname='Bhat]{Atul Bhat}
\affiliation{Manipal Centre for Natural Sciences, Manipal Academy of Higher Education, Manipal, Karnataka 576104, India}
\email[]{atul.mcnsmpl2023@learner.manipal.edu}  

\author[orcid=0000-0002-7276-4670,gname=sreejith, sname='Padinhatteeri']{Sreejith Padinhatteeri} 
\affiliation{Manipal Centre for Natural Sciences, Manipal Academy of Higher Education, Manipal, Karnataka 576104, India}
\email['show']{sreejith.p@manipal.edu}

\author[orcid=0000-0003-4908-6186]{J.M. Borrero}
\affiliation{Institut für Sonnenphysik (KIS), Georges-Köhler-Allee 401A, 79110, Freiburg, Germany}
\email[]{borrero@leibniz-kis.de} 

\author[0000-0003-0585-7030]{Jayant Joshi}
\affiliation{Indian Institute of Astrophysics, Koramangala, Bangalore 560034, India}
\email[]{jayant.joshi@iiap.res.in}

\correspondingauthor{Sreejith Padinhatteeri}
\email{sreejith.p@manipal.edu}

\begin{abstract}
Light bridges are commonly observed in active regions and are interpreted as signatures of magnetoconvective processes in sunspots. Several studies have attempted to classify them in the past based on their morphological characteristics. Recent observations have revealed new dynamical properties of light bridges, including their signatures in the upper solar atmosphere, particularly in the chromosphere, and their coupling with coronal features. In this study, we observed two cases of rare and unusual dynamics as light bridges evolve. Using data from the Solar Dynamics Observatory, the evolution of the light bridges is analysed, and the results are reported here. Based on our findings, we propose that the unique movements of the light bridges in the observed sunspot and earlier studies could be an apparent view of the umbral core dynamics. Investigation into these dynamics through signatures in the higher atmosphere reveals a clear coupling to coronal loops and their dynamics. 
 
\end{abstract}


\section{Introduction} 

    \begin{figure*}[t]
        \begin{interactive}{animation}{anim_13590_frames.mp4}
            \centering
            \includegraphics[width=1\linewidth]{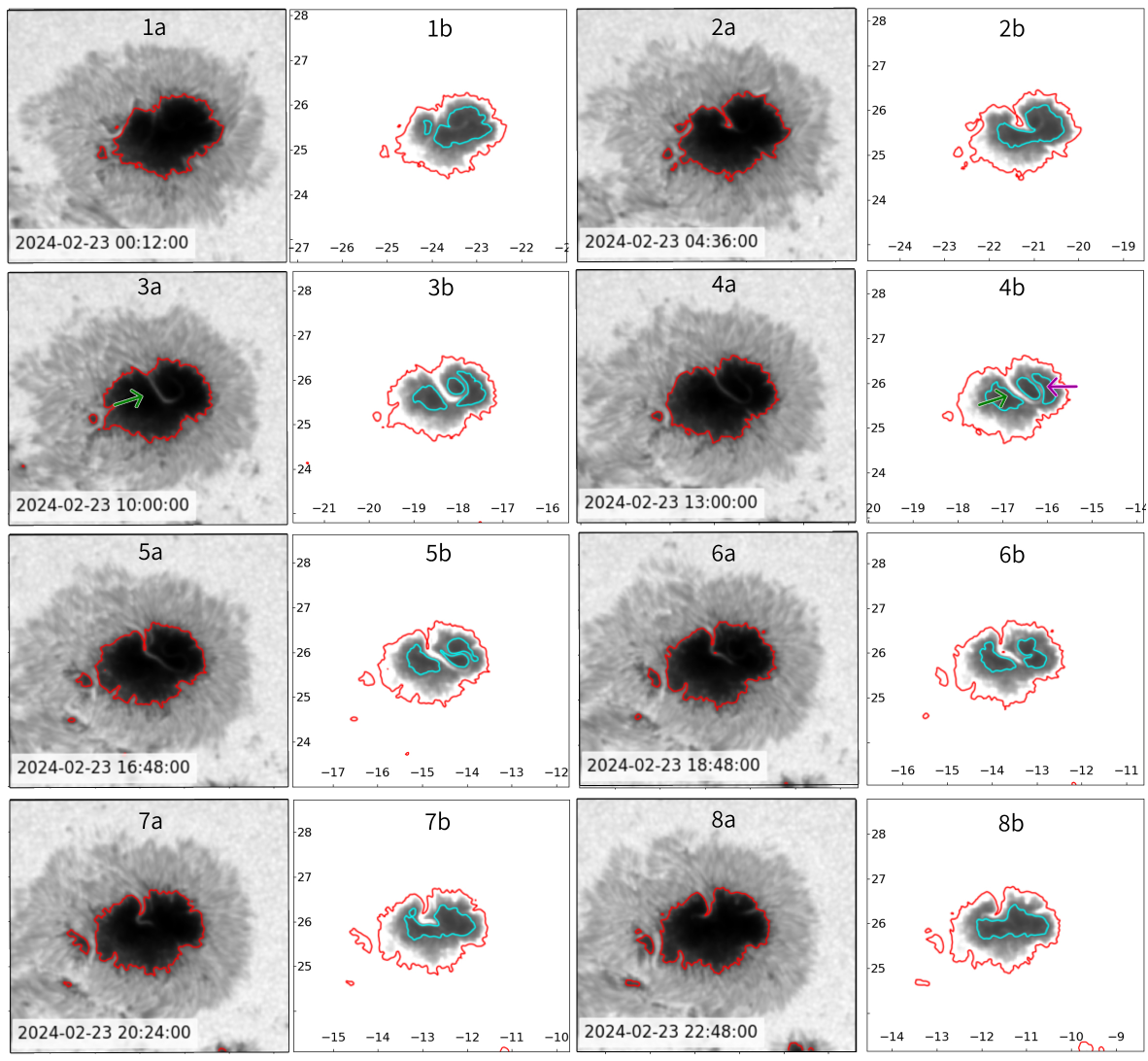}
            
        \end{interactive}
        \caption{Normal contrast (\textit{a frames}) and the corresponding enhanced contrast (\textit{b frames}) images of the sunspot in AR 13590. Each tick on the x and y axes represents latitude and longitude in Stonyhurst coordinates, respectively. The coordinates of \textit{a frames} are the same as the corresponding \textit{b frames}. The timestamps apply to each frame-pair. These panels show the $S$-light bridge visibly appearing and evolving, before disappearing within a span of 24 hours. The actual $S$ shape of the light bridge is most clearly visible in frame \textit{5b}. The red contours mark the umbral boundary, identified by iso-countours at $0.4I_\text{QS}$. The cyan countours at $0.09I_\text{QS}$. In frames 3a and 4b the green arrow points to the light bridge \textit{arm} and the magenta arrow in 4b points to the faint \textit{tail}. The Animation shows normal contrast and enhance contrast frames, side-by-side, of time-evolution of the light bridges for the duration of 24 hours on 23 Feb 2024. }
        \label{fig:cont_enhanced}
    \end{figure*}

    Light Bridges (LB) are bright elongated features that split the sunspot umbra into smaller regions \citep{Bray1964}. Previous works have identified various types and properties of light bridges, such as those studied by \citet{Sobotka1997} where LBs do not break the umbra into smaller pieces, or by \cite{Vazquez1973} who observed LBs as penumbral-intensity filaments. A detailed analysis of LBs by \citet{Leka1997, Livingston1991, Beckers1969, Lites1991} established properties such as a drop in the magnetic field and an increase in the inclination of the magnetic field across the width of the LB. Studies have also classified LBs into various categories, based on their morphology and behavior. Light bridges can be granular, that split the umbra, or arise as umbrae coalesce \citep{Bray1964, Garcia1987, Leka1997, Lagg2014}, or `faint' penumbral brightness filaments that protrude into the umbra \citep{Vazquez1973, Muller1979, Livingston1991, Katsukawa2007, Louis2008, Bharati2015}. In particular, \citet{Muller1979} classified them as (a) photospheric LBs that have granules but smaller than typical quiet sun granules; (b) penumbral light bridges that are elongated filaments with same brightness as the penumbra and also line up with the filaments in the penumbra; and (c) umbral light bridges that are streamers with dots, where these dots are brighter than umbral dots. 
    
    Light bridges are mostly associated as photospheric features because most of the details that we observe in them are generally visible at photospheric heights. Multiple observations of LBs have been made in the chromosphere \citep[for a review see][]{Solanki2003}. Plasma ejections with jet-like features have also been observed in H$\alpha$ \citep{Asai2001} and Ca II H (3968 {\AA}) images \citep{Shimizu2009} in the chromosphere.  \citet{Bharati2007} describe these plasma ejections as being caused by magnetic reconnection between the low chromosphere and the upper photosphere. This magnetic reconnection may lead to the injection of dense plasma into the chromosphere. \citet{Korobova1966} has also described an additional type of LB, referring to them as faint narrow streamers that are the projection of the chromospheric filaments. These results established that LBs can result in photospheric-chromospheric coupling. \citet{Feng2020} and \citet{Miao2021} have also observed instances where coronal loop and light bridge dynamics show correlations. \\

    \citet{Schlichenmaier2016} have found LBs that are neither granular nor filamentary. They report on a new type of LB with central dark lanes and transverse bright and dark lanes that are not connected to the central dark lanes (see their Figs. 9 and 13). These LBs do not form around coalescing or decaying sunspots. Another type of unconventional LBs come from the studies by \citet{Kleint2013} and \citet{Guglielmino2017}, who reported features that are visually similar to LBs, but since these filament-like structures do not break the umbra, they coined the term `umbral filaments'. These umbral filaments were of penumbral intensity, protruded into the umbra from the penumbra, and had hook-like tips. Although they appear as regular LBs, the drop in values of magnetic field across these filaments was smaller compared to conventional LBs. The lifetime of these umbral filaments was also observed to be quite short compared to that of conventional granular LBs. Both observations mention `hook'-like structures observed within the filaments. They also report `a flow opposite to the Evershed Flow' within these umbral filaments, which resembles Counter-Evershed Flow \citep[CEF;][]{Louis2014, Siu-Tapia2018}. A recent study by \citet{Duran2021} revealed that CEFs are commonly observed around light bridges. \\
    
    \cite{Xing2024} have identified footpoints of CME flux ropes near `sunspot scars' that manifest as light bridges in the photosphere. They studied the magnetic field components of a sunspot scar and find that the total magnetic field intensity drops across it, and so does the vertical component of the magnetic field. It was also observed that the horizontal component of the magnetic field rises across the sunspot scar (see their Figs. 4 and 5). \\
    
    In this paper, we study the aforementioned umbral filaments type of LBs that do not break the umbra. In particular, we focus on their evolution and their peculiar transverse movement within the umbra. If LBs are indeed field-free protrusions into the strong vertical umbral magnetic field \citep{Leka1997}, then they are not expected to move transversely crossing through the vertical magnetic field lines. Although few LBs have previously been reported as a consequence of umbral coalescence, to the best of our knowledge, no study has systematically examined their apparent transverse motions, particularly in cases where umbrae do not merge and the LB does not disrupt the umbral structure.  \\
    
    We have also investigated whether the coupling of the umbral magnetic field lines with the chromosphere and the corona is contributing to the transverse motion. In this paper, we try to answer what could be the driving mechanism that causes the transverse motion of filament-type LBs. Section \ref{section:observations} explains the details of our observation, the data sets used, and the preliminary analysis. Section \ref{section:results} consists of the various results we have obtained from our analysis of LBs in the photosphere, chromosphere, and corona. In section \ref{section:discussion} we summarize our findings and present our conclusions. \\

\section{Observations \& Data Analysis}
\label{section:observations}

    In this paper, we used data from the Solar Dynamic Observatory \citep[SDO;][]{Pesnell2012} for the study of LBs. The intensity and magnetic parameters are obtained from The Helioseismic and Magnetic Imager \citep[HMI;][]{Scherrer2012, Schou2012}. We used HMI Space-Weather Active Region Patches \citep[SHARPS;][]{Hoeksema2014} data with 720s cadence. To avoid projection effects, we used the data from SHARP Cylindrial Equal Area (CEA) series. We also used the Atmospheric Imaging Assembly \citep[AIA;][]{Lemen2012} for the chromospheric (1700\r{A} and 1600\r{A}) as well as coronal (171\r{A}) data of the LBs and associated coronal features. \\

    We report multi-height observations of an LB from NOAA AR 13590 (NOAA Active Regions referred to as AR for the rest of the paper), which appeared on 23 February 2024. We refer to this LB as the $S$-light bridge. We observed an apparent transverse motion of the $S$-light bridge which caught our interest in investigating it further. A similar LB within the same sunspot appeared on the previous day, 22 February. We call this the $\Omega$-light bridge, based on the shapes they evolve into.  \\

    \begin{figure*}[!t]
        \begin{interactive}{animation}{anim_13590_22_Frames.mp4}
            \centering
            \includegraphics[width=\textwidth]{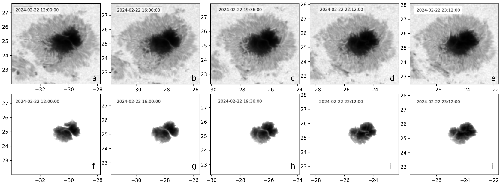}
        \end{interactive}
        \caption{The $\Omega$-light bridge observed on 22 Feb 2024. The top row shows the normal contrast images from HMI and the bottom row shows the same images with enhanced contrast. The light bridge forming the $\Omega$ shape shape is visible in \textit{frame i}.The Animation shows normal contrast and enhance contrast frames, side-by-side, of time-evolution of the light bridges for the duration of 24 hours on 22 Feb 2024.}
        \label{fig:13590-22-frames}
        \vspace{0pt}
    \end{figure*}
     
    The host sunspot in AR 13590 appeared at the eastern limb on 18 February 2024 and moved across the solar disk, at an approximate latitude of $25^0$ N from the solar equator, until the 3 March 2024, when it disappeared at the west limb. The $S$-light bridge was observed on 23 February 2024 between 00:12 UT and 22:00 UT, and the $\Omega$-light bridge was observed on the previous day on 22 February 2024 between 10:00 UT and 23:00 UT at a $\mu = ~0.8$. Both light bridges were very faint. The contrast had to be enhanced for them to be visible as they were almost invisible at normal dynamic range of intensity (see panels marked (b) in Fig.~\ref{fig:cont_enhanced} and bottom row in Fig.~\ref{fig:13590-22-frames}).  \\

    The Dopplergrams from the HMI SHARP data were corrected for LOS velocity by subtracting the Quiet Sun average velocity for each Dopplergram data. Generally $40\%$ of Quiet-Sun mean intensity is used as a threshold for umbral intensity $I_c = 0.4I_\text{QS}$ \citep{Leka1997}. However, we noticed that using the the aforementioned threshold makes both LBs indistinguishable from the umbra. Therefore, in order to show the umbra by excluding the LB, we use $0.09I_\text{QS}$ as the threshold for LB excluded umbra. In Fig.~\ref{fig:cont_enhanced} we have plotted the sunspot of AR 13590 with $I$ < $0.4I_\text{QS}$ in red to mark the umbra in all frames. In the contrast enhanced frames ($b frames$) we also mark the LB-Excluded-Umbra with $I$ < $0.09I_\text{QS}$ in cyan. Unless otherwise specified, the same isocontours at $0.09I_\text{QS}$ have been used for all plots and images.   \\

    To characterize the LB motion we also defined umbral core regions (see \S~\ref{section:umbralcore}) with magnetic fields total magnetic field $B > 2 $kG and intensity $ 8000$ DNs $ \sim I < 0.07I_\text{QS}$. This helps us identify regions to study the apparent transverse motion of the LBs. We do this by calculating the weighted centroids of the umbral core regions around the LBs. We studied the motion of these centroids and the changes in their morphology to understand whether and how they govern the motion of LBs.\\

\section{Results}
\label{section:results}

    We observed two cases of apparent transverse motions of LBs within the umbrae of AR 13590. We observed the ARs in different wavelengths, corresponding to the plasma properties at different atmospheric heights of the Sun. \S \ref{section:photosphereobs}, \ref{section:chromosphereobs}, \ref{section:coronalobs}, explain the results of the observations in Photosphere, Chromosphere, and Corona, respectively, for the LBs. The last subsection \S \ref{section:umbralcore} explains the results of the analysis of umbral core dynamics. \\
    
    \begin{figure*}
        \centering
        \includegraphics[width=0.9\linewidth]{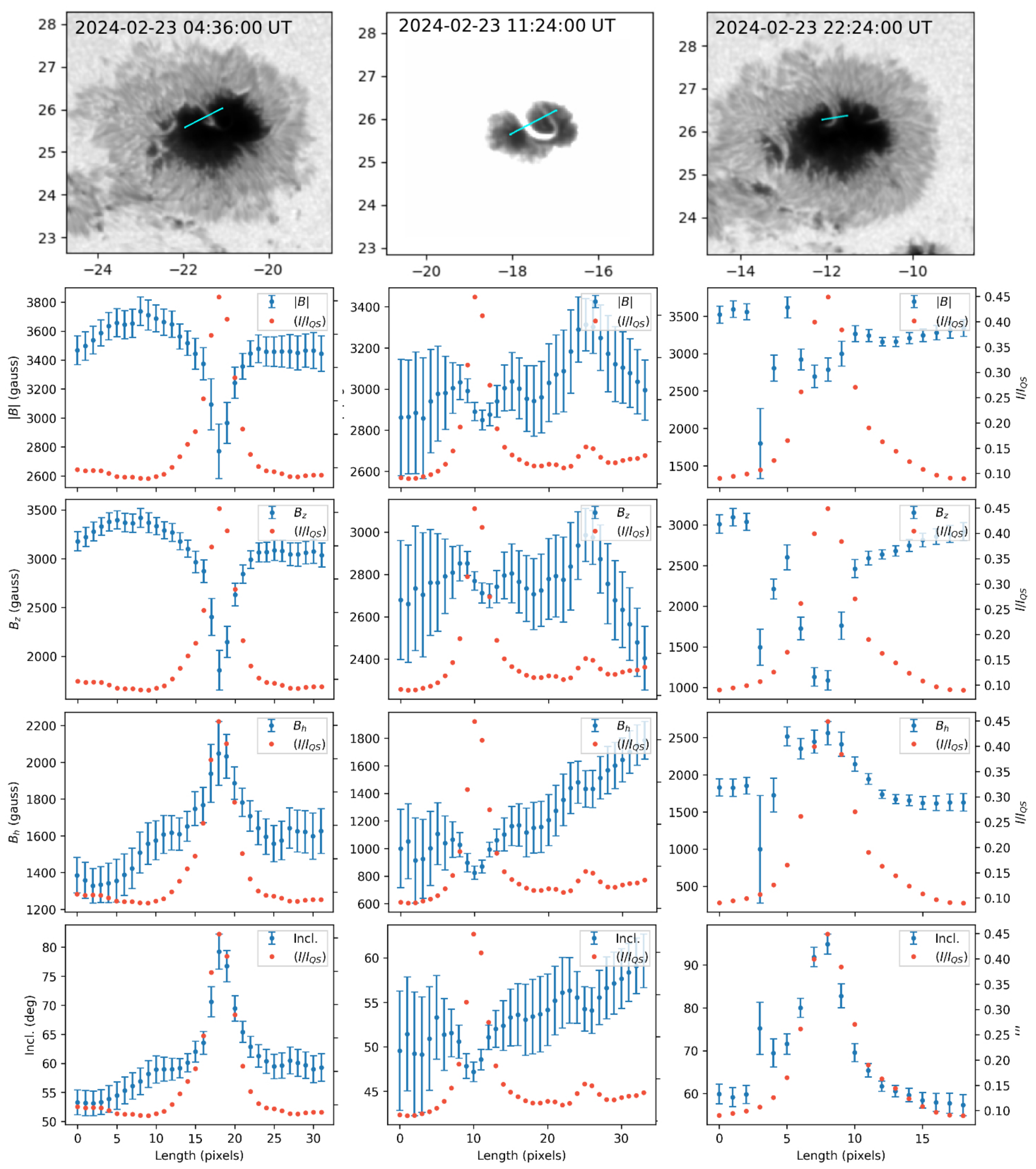}
        \caption{Continuum intensity AR 13590 (top row) showing a cut across the light bridge (cyan line); the magnetic field intensity, vertical component of the magnetic field, horizontal component of the magnetic field and inclination with respect to LOS (subsequent rows; blue) plotted along line across the light bridge at various timestamps. The ticks in the top row are latitude and longitude in Stonyhurst coordinates. The vertical axes on the right of each plot (labeled in the last columns) show the normalized intensity $I/I_\text{QS}$ (marked by red dots) in each plot. The continuum intensity frames in the first and last column show normal light bridge behavior in the absence of the \textit{tail}. Middle frame shows high contrast intensity image to make the faint \textit{tail} visible. For this timestamp (11:24 UT), the drop in the magnetic field intensity along the light bridge is not consistent at the two points where the line meets the light bridge. The horizontal component of the magnetic field does drop at both the points, so does the inclination. }
        \label{fig:inclination-field-plot}
    \end{figure*}

    \subsection{Photospheric observations}
    \label{section:photosphereobs}

        Figure~\ref{fig:cont_enhanced} shows the evolution of a $S$-light bridge into the umbra of AR 13590. Frame numbers 1a-8a show time series frames in normal contrast, and 1b-8b show the corresponding high-contrast images to show the LB apparent motions. \\

        Unlike most observed LBs, the $S$-light bridge on AR 13590 did not form due to the coalescence of sunspots, and it did not break the umbra.  Moreover, no granular structure was observed throughout its evolution at HMI resolution. This LB was studied for its peculiar behavior where it behaved like an umbral filament - where it reached two points on the umbra-penumbra boundary, but did not break it into two similar to that observed \citet{Kleint2013}, and also behaved like a normal LB - with lower magnetic field field strength and higher inclination compared to the surrounding umbra \citep{Leka1997}. Further, as this LB protruded and reached the opposite end of the umbra, it retracted with a `whip'-like transverse motion. This retraction that was transverse in nature, along with the non-breaking of the umbra seemed abnormal. Further, as we enhance the contrast of the intensity image, we noticed a faint \textit{tail} as an extension to the $S$-light bridge which was not visible at normal contrast (see arrows in Fig.~\ref{fig:cont_enhanced} [two movies of frames a (normal contrast) and b (high contrast) frames are provided as supplementary material]).\\

        \begin{figure}
            \centering
            \includegraphics[width=1\linewidth]{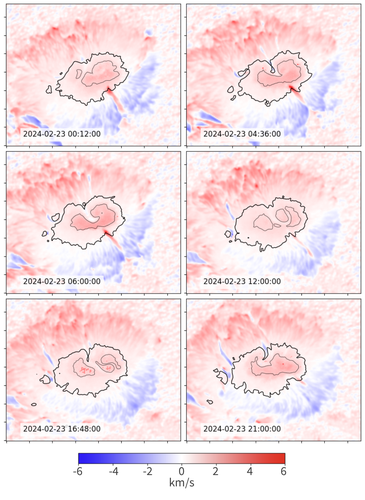}
            \caption{$V_{LOS}$ for the sunspot with the $S$-light bridge. Contours are plotted for intensity at $0.4I_\text{QS}$ (outer) and $0.09I_\text{QS}$ (inner) plotted over it. CEFs are observed along the \textit{arm} of the $S$-light bridge as it initially extends and when it recedes back. No CEFs are observed along the \textit{tail} of the light bridge. The large values in the core-umbra (marked by the gray contour) may be due to saturation effects and other factors.}
            \label{fig:dopplergrams}
        \end{figure}

        The following sequence of events was observed over a span of 24 hours. The $S$-light bridge first forms as an \textit{arm} protruding from the penumbra into the umbra around 00:12 UT on 23 Feb. Around 04:00 UT the fainter \textit{tail} of the LB appears as a curved extension to the LB. When the \textit{arm} of the LB reaches the opposite side of the umbra (in frame \textit{5b}), the \textit{tail} forms the $S$-shape. As time progresses, the \textit{tail} grows into a larger $S$, while the \textit{arm} also grows slightly, forming an `elbow'-like bend along its length. At 18:48 UT, the \textit{arm} and \textit{tail} break apart, and the \textit{arm} undergoes the whip-like motion. During this time, the tail slowly disappears.

        \begin{figure}
            \centering
            \includegraphics[width=1\linewidth]{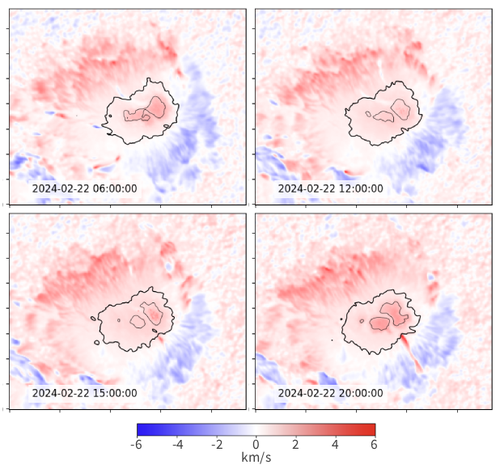}
            \caption{ $V_{LOS}$ for the sunspot with $\Omega$-light bridge. Contours same as Fig.~\ref{fig:dopplergrams}. CEFs are observed for the two light bridges individually, which appear and disappear independently. Each x tick marks 50 pix and the y ticks mark 25 pix.}
            \label{fig:dopplergrams_22}
        \end{figure}
        
        A second case of similar dynamics was observed in the case of the $\Omega$-light bridge, where initially there were two individual LBs, one protruding into the umbra from the northern side and the other from the southern side of the umbra-penumbra boundary. Both LBs were very faint and clearly visible after enhancing the contrast of the continuum images. Frames a and f in Fig.~\ref{fig:13590-22-frames} show the two separate LBs in normal and enhanced contrast, respectively. The protruding ends of the two LBs connect and merge into a single LB at 16:00 UT (see frames g through i in Fig.~\ref{fig:13590-22-frames}). Thereafter, beginning at 16:12 UT, the LB's central region appears to move transversely until 22:24 UT, forming the distinct $\Omega$-shape at 22:12 UT (frame i in the aforementioned image). The LB begins to disintegrate and disappears at 23:12 UT. Unlike the $S$-light bridge, the $\Omega$-light bridge undergoes these changes in about 12 hours. The total magnetic field intensity and the magnetic field components were measured across the LB and were found to have similar patterns. \\

        From the plots of the magnetic field components and inclination across the $S$-light bridge (see Fig.~\ref{fig:inclination-field-plot}), we note that in the absence of the \textit{tail}, the \textit{arm} shows an increase in the horizontal component of the magnetic field across the LB and a drop in the vertical component and the total magnetic field intensity. When the \textit{tail} appears (middle row in the aforementioned figure), the horizontal magnetic field component drops across the \textit{arm} and the \textit{tail}. The vertical magnetic and the total magnetic field intensity drops across the \textit{arm} and show a rise across the \textit{tail}. The inclinations of the magnetic field vector rise across the LB \textit{arm} before the formation of the \textit{tail}. When the \textit{tail} has formed, the inclination across both the \textit{arm} and \textit{tail} shows a drop. \\ 
        
        We corrected the Dopplergrams ($v_{LOS}$) by measuring the average velocity for a large Quiet-Sun region within the SHARP patch and subtracting the data with value, so that the average line-of-sight velocity in the Quiet Sun must be zero. The quiet sun region selected for correction spanned 700 pixels across the top 70 rows of the SHARP data. It must be noted that the velocities within the core-umbra (marked by the inner gray contour in Fig.~\ref{fig:dopplergrams}) show downflow, which is unusual. We also tried taking the quiet-sun region for correction as a small region (100x100 pixels) along the same longitude as the sunspot. However, no changes were observed. We also measured the magnetic field intensity within these contours. We noticed large variations ($\Delta|B| > 2000$G) during the LB evolution. These irregular values could be due to factors like saturation effect. We therefore ignore the velocities and the magnetic field variation within the umbra for this study. \\
        
        The Dopplergram for the $S$-light bridge (see Fig.~\ref{fig:dopplergrams}) also shows flows that appear as CEFs. We measure the CEFs above the \textit{arm} with an inward flow $\sim1.5$ $kms^{-1}$ while the outward velocity in the surrounding penumbra was approx. between $0.4$ and $1.1$ $kms^{-1}$ at $\mu = 0.84$ (data from 23 Feb, 12:00 UT). The CEFs in the lower half of the umbra-penumbra boundary go as high as $\sim 5$ $kms^{-1}$ during the LB evolution (right image in the first row of the same figure), before the \textit{tail} forms.  We also observe that the CEFs disappear as the \textit{tail} touches the edge of the penumbra, and the inward velocity returns as the \textit{arm} separates and executes the whip-like motion. \\
        
        CEFs were also observed in the case of the $\Omega$-light bridge, where individual CEFs appeared for the two LBs that joined to form the $\Omega$-light bridge (see Fig.~\ref{fig:dopplergrams_22}). The LB that formed near the upper edge of the umbra-penumbra boundary around 5:00 UT (top-left frame) and disappeared just before the appearance of the lower LB at 12:00 UT (top-right frame). However, a second CEF emerged near the umbra-penumbra boundary where the second LB appeared (bottom-left frame) and the CEF extended towards the outer edge of the penumbra (bottom-right frame) and remained there, reaching a max. velocity (LOS) of $\sim 4 $ $kms^{-1}$, before it disappeared.\\
    
        \begin{figure}
            \centering
            \includegraphics[width=1\linewidth]{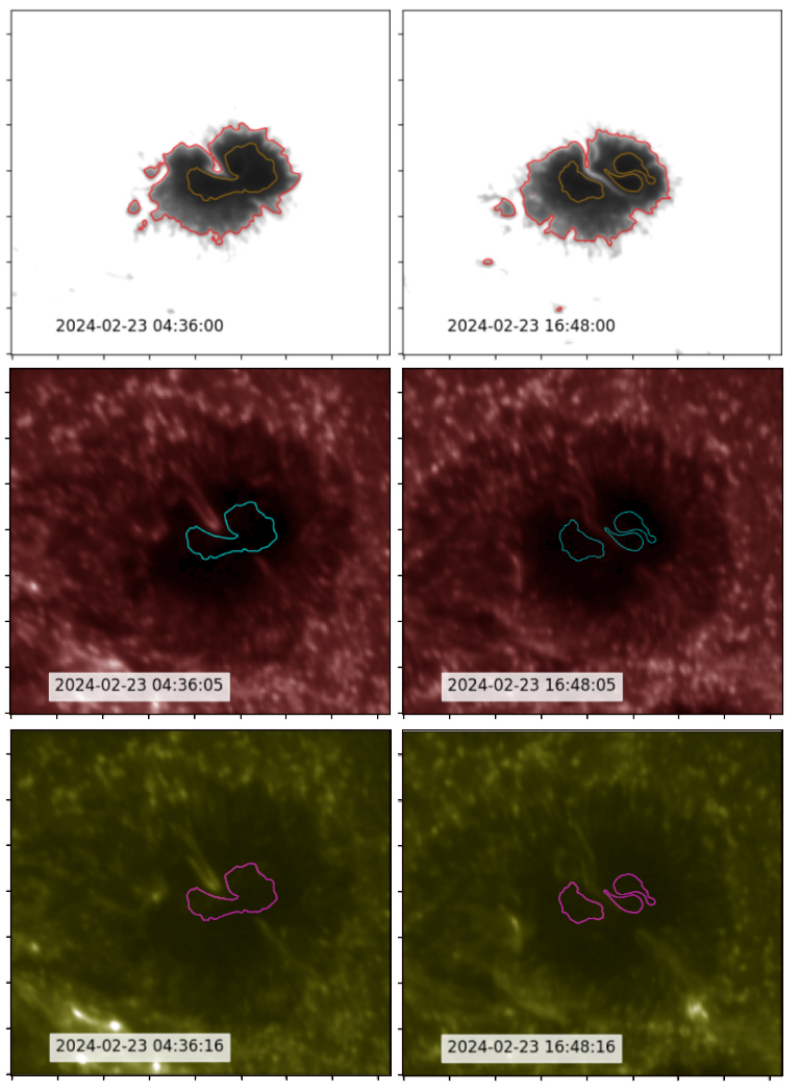}
            \caption{The $S$-light bridge of AR 13590 in photosphere (HMI enhanced contrast - First row) and the chromosphere (AIA 1700\r{A} - middle row and AIA 1600\r{A} - bottom row). Left columns (4:36 UT) show brightening in the AIA frames along the \textit{arm} of the LB. The right frames (16:48 UT) show both the \textit{arm} and the \textit{tail} of the LB, through contours marked for $0.09I_\text{QS}$ in the AIA frames The red contour in the HMI frames mark the umbral boundary. The \textit{tail} of LB is not visible in the AIA frames. Each tick marks 25 pixels.}
            \label{fig:chromosphere-plots}
        \end{figure}

        \begin{figure}
            \centering
            \includegraphics[width=1\linewidth]{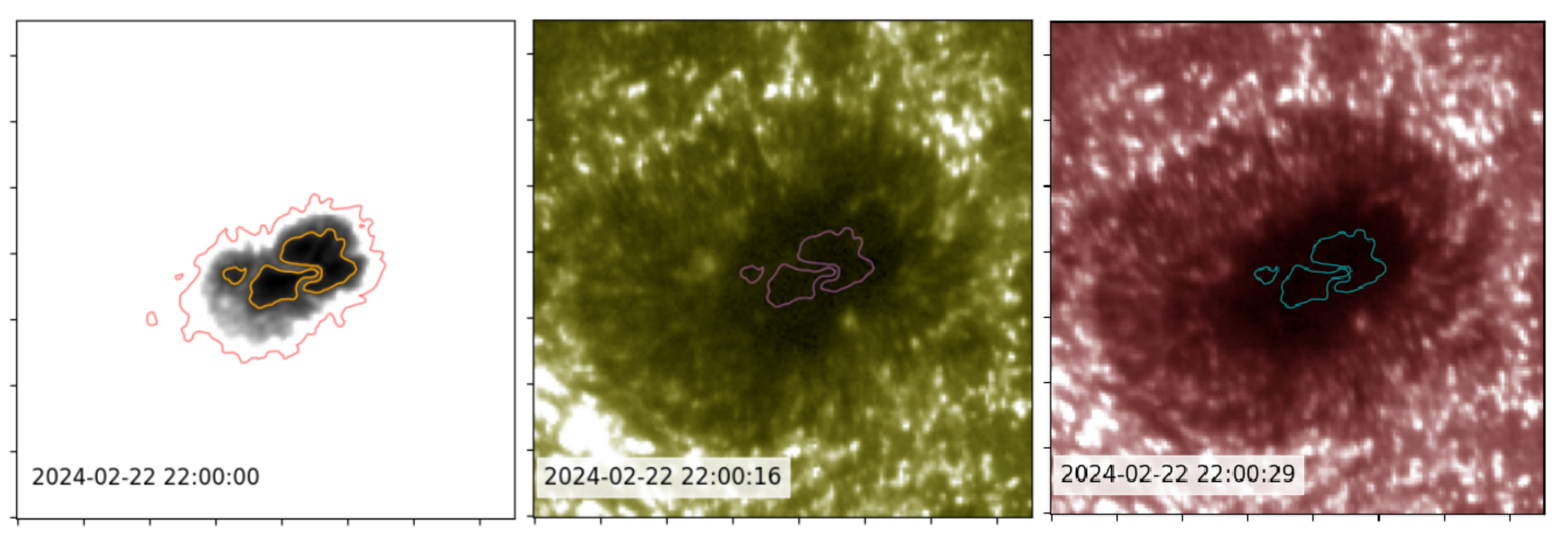}
            \caption{The $\Omega$-light bridge of AR 13590 in photosphere (HMI enhanced contrast - left) and the chromosphere (AIA 1600\r{A} - middle and AIA 1700\r{A} - right). The contours are marked for $0.09I_\text{QS}$ in the AIA frames and the red contour in the HMI frames mark the umbral boundary. The light bridge is not visible in the AIA frames. Each tick marks 25 pixels.}
            \label{fig:chromosphere-plots_22}
        \end{figure}
        
    \subsection{Chromospheric observations} \label{section:chromosphereobs}

        \begin{figure*}[!ht]
            \centering
            \includegraphics[width=1\linewidth]{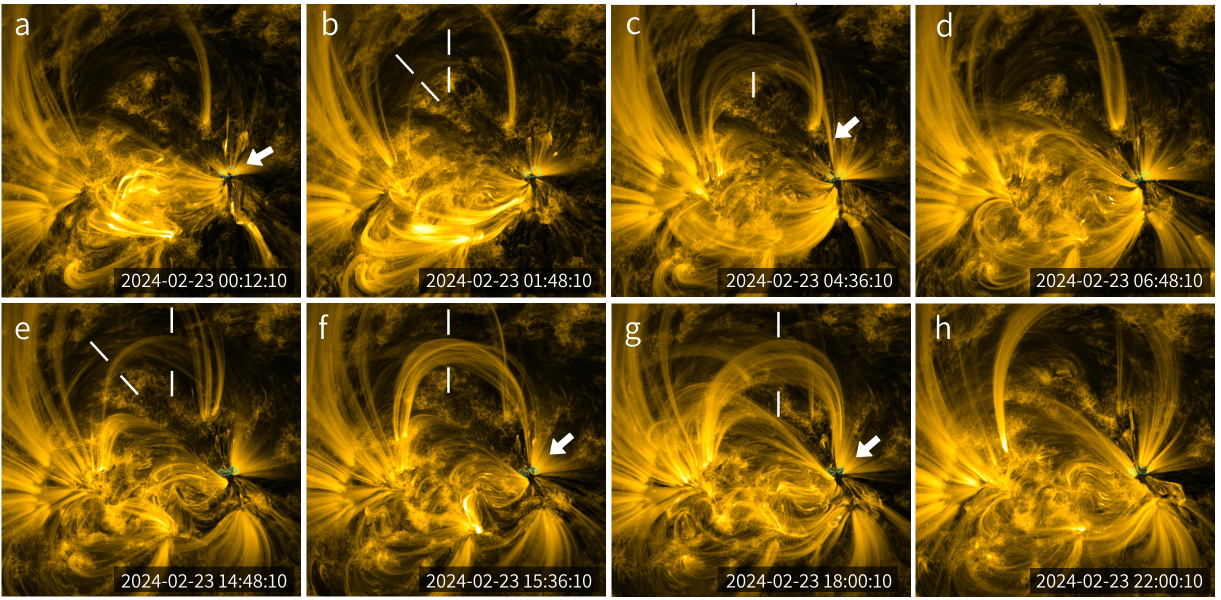}
            \caption{A 420"$\times$420" FOV of the AR 13590 in AIA 171\r{A} at various timestamps on 23 Feb 2024. The umbra with the $S$-light bridge is marked with cyan contour (marked by white arrow in frame 'a'). The white lines with central gaps indicate the coronal loops (visible between the gaps). A faint coronal loop forms (frame 'b' \& 'c') with its footpoints (marked by arrow in frame 'c') in the sunspot. Another loop (frame 'e' through 'g') appears which temporally correlates with the LB dynamics. The coronal loops in these frames grow, wiggle around and disappear (frame 'h').}
            \label{fig:coronal-loop-large}
            \vspace{-2pt}
        \end{figure*}

       \begin{figure}[!b]
            \centering
            \includegraphics[width=1\linewidth]{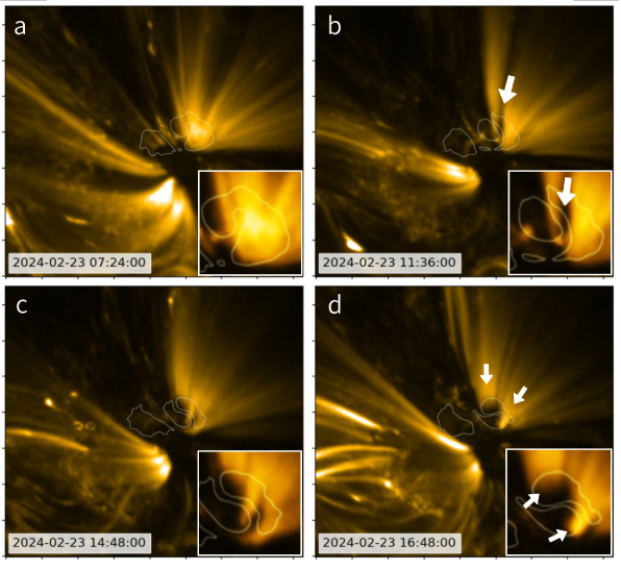}
            \caption{The sunspot from Fig.~\ref{fig:coronal-loop-large} zoomed-in to 100"$\times$100" FOV. The coronal loop footpoints within the umbra (faint white contour for $0.09I_\text{QS}$). The inset shows a further zoomed-in view. a) footpoint on the right appears as a single footpoint. b) as the \textit{tail} extends, a gap within the footpoints emerges - marked by white arrows. c) The curved separation of footpoints can be traced along the light bridge \textit{tail}. d) the two footpoints on either side of the light bridge are distinctly visible - marked by white arrows.}
            \label{fig:coronal-loop-small}
        \end{figure}
        
        To better understand the nature of the LBs and the observed transverse motion in both, we studied this LB in the AIA 1700{\r{A}}  and 1600{\AA} wavelengths. The $S$-light bridge initially shows brightening in the chromospheric wavelengths (see Fig.~\ref{fig:chromosphere-plots}), compared to the surrounding regions, as has been observed by others \citep{Leka1997, Duran2021, Kleint2013}. Although the \textit{arm} initially shows brightening, the \textit{tail} shows no signatures in the AIA frames. Since only the \textit{arm} is visible in the chromospheric layers, we can infer that the tail is a low-lying (i.e. mainly photospheric) feature. A similar case was found for the $\Omega$-light bridge (see Fig.~\ref{fig:chromosphere-plots_22} where the entire LB was not visible in the AIA UV frames, suggesting that the $\Omega$-light bridge was photospheric only.  \\

    \subsection{Coronal Dynamics}
    \label{section:coronalobs}

        To investigate the apparent transverse motion of the \textit{tail} of the $S$-light bridge and the $\Omega$-light bridge, we looked into the higher atmosphere using the AIA 171\r{A} images. We observed several corresponding events along the timeline of the LB evolution. The images were de-convolved as described by \citep{Hofmeister2021}. \\

        \begin{figure*}[!t]
            \centering
            \includegraphics[width=\textwidth]{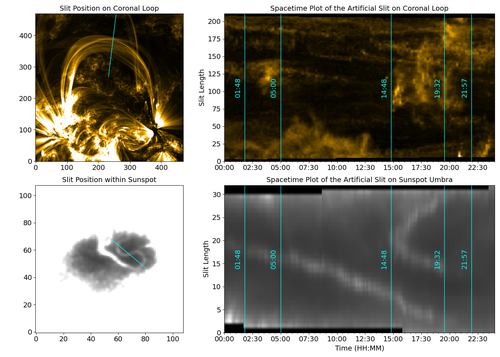}
            \caption{SpaceTime (st) plots of artificial slits on coronal loops (top) and the light bridge (bottom) in AR 13590 on 23 Feb 2024. The left square plots show the position of the slit on the features and the plots on the right are the slits over time. The occurrence of loop shown in Fig.~\ref{fig:coronal-loop-large} frames b and c,  and the appearance of second loop in Fig.~\ref{fig:coronal-loop-large} frames e through g, are marked in both st plots. The coronal loop appearance and disappearance corresponds to the formation of the $S$ shape of the light bridge.}
            \label{fig:stplot}
        \end{figure*}
        
        In the AIA 171\r{A} images for 23 Feb, we distinctly notice, at 01:48 UT, the appearance of a large coronal loop anchored at the second-largest sunspot (see Fig.~\ref{fig:coronal-loop-large}), where the $S$-light bridge exists. It disappears soon after the chromospheric brightening at 04:36 UT (frame d). Several such coronal loops were observed with their footpoints within the sunspot. The formation of coronal loops around an LB formation has been previously observed by \citet{Miao2021}. Later, at 14:47 UT, a very similar large loop reappears, this time more quickly, and interacts with the observed sunspot. This loop disappears at 22:00 UT. \\

        As these timings correlate with the $S$-light bridge observations, we plotted intensities over time across artificial slits that cut the coronal loops and the LB as shown in Fig.~\ref{fig:stplot}. The slit across the sunspot umbra is plotted for a contrast-enhanced image. The appearance of the first loop is clearly visible through the brightening between 01:48 and 05:00 UT (frames b and c in Fig~\ref{fig:coronal-loop-large}). The appearance of the second larger loop occurs from 14:48 to 22:00 UT across the slit. Even at enhanced brightness, the LB appears until 21:57 UT, and immediately afterwards it seems to touch the umbra-penumbra boundary as it becomes indistinguishable from the surrounding umbra. The coronal loop begins to fade around this time and it disappears at 21:57 UT. Between 19:32 and 21:57 UT, the loop brightening is present only in the upper portion of the slit. The bright points on the rest of the slit is due to the presence of other loops and the background. Upon closer observation in the vicinity of the LB, the AIA images also reveal the footpoints of these coronal loops within the umbra (see Fig.~\ref{fig:coronal-loop-small}). \\

        In addition, as we look closely at the footpoints of the coronal loop around the $S$-light bridge, over-plotted with the contour of the umbra at $0.09 I_\text{QS}$, the footpoints of various coronal loops become clearly visible (see Fig.~\ref{fig:coronal-loop-small}). As the LB evolved, at 11:36 (frame b) the bright footpoint can be seen with a gap forming within. We note the \textit{tail} - through the umbral contour, passing through this gap, thus allowing us to assume that the coronal-loop footpoints exist on either side of the \textit{tail}. At 14:48 UT (frame c) this becomes evident, as the contour of the \textit{tail} appears to curl around the footpoint. The two footpoints moving around become clearly visible at 16:48 UT (frame d). \\

        \begin{figure}
            \centering
            \includegraphics[width=1\linewidth]{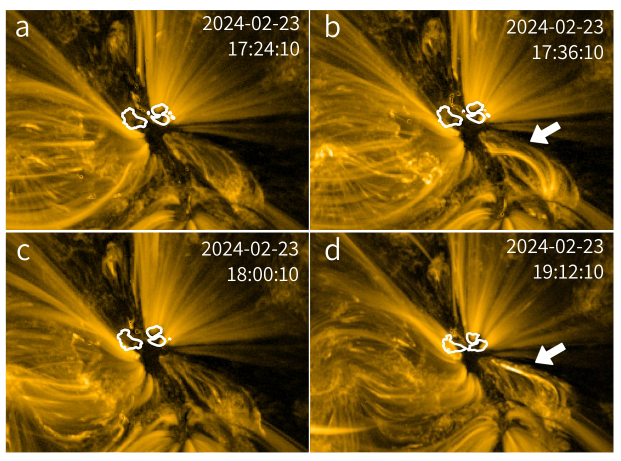}
            \caption{135"$\times$100" frames from AIA 171\r{A}  centered at the umbra (cyan contour for $0.09 I_\text{QS}$) for $S$-light bridge. The small loops correspond to LB events discussed in last subsection in \S~\ref{section:discussion}. Frames a and c show the pre-loop stages, while \textit{frame b} shows the small coronal loop (pointed by the arrow), which occurs as the LB \textit{arm} and \textit{tail} break and \textit{c} frame d shows another such loop when the \textit{tail} disappears.}
            \label{fig:coronal-loop-cea}
        \end{figure}

        Closer inspection of the LB evolution revealed additional small-scale coronal loops (see Fig.~\ref{fig:coronal-loop-cea}). For instance, at 17:36 UT a small loop quickly forms and disappears as the LB \textit{tail} and \textit{arm} break. A very similar loop shows up at 19:12 UT, just before the \textit{tail} of the LB disappears. \\

        \begin{figure}
            \centering
            \includegraphics[width=1\linewidth]{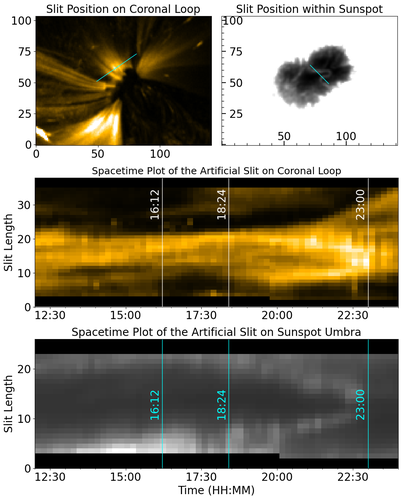}
            \caption{SpaceTime (st) plots of artificial slits across coronal loops (middle) and the $\Omega$ light bridge (bottom) in AR 13590 on 22 Feb 2024. The square plots on the top show the position of the slit (slit from left to right) on the features and the plots on the right are the slits over time. The expansion of the loop shown in Fig.~\ref{fig:coronal-footpoint-22} is visible in the AIA st plot and the corresponding timestamps mark the appearance of the light bridge within the slit.}
            \label{fig:stplot22}
        \end{figure}

        \begin{figure*}[t]
            \centering
            \includegraphics[width=1\linewidth]{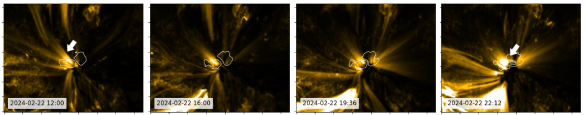}
            \caption{Frames from AIA 171\r{A} at during the $\Omega$-light bridge evolution on 22 Feb 2024. The umbral core regions are contoured in white. The coronal loop is marked with the white arrow in first and last frames. The coronal loop footpoint show lateral expansion in each frame.}
            \label{fig:coronal-footpoint-22}
        \end{figure*}
        
        We also noticed a bright loop with footpoint in the sunspot for 22 February, when the $\Omega$-light bridge appeared. We observe that the coronal loop footpoint (marked by arrows in Fig.~\ref{fig:coronal-footpoint-22}) grows broad over time, which is evident from the space time plot in Fig.~\ref{fig:stplot22} with the slit drawn across the footpoint. The loop begins to broaden around 16:23 UT, around the same time the LB undergoes the apparent transverse motion, which is visible in the plot ( lower row of the same Fig.). The LB `folds' until its disappearance at 23:00 UT. \\

    \subsection{Umbral Core Dynamics}
    \label{section:umbralcore}
    
        \begin{figure*}
            \centering
            \includegraphics[width=1\linewidth]{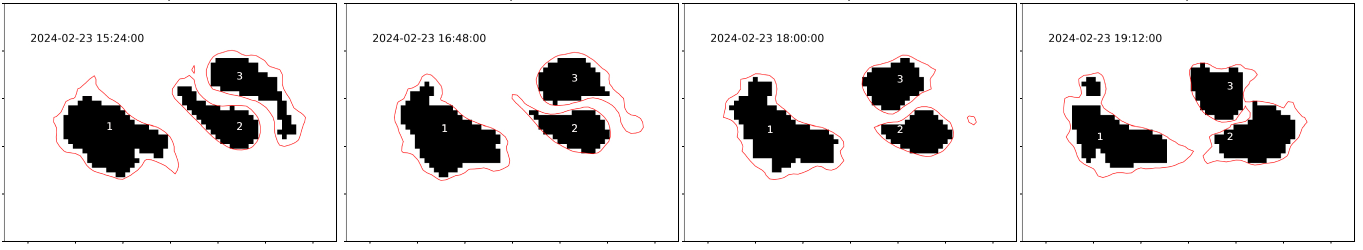}
            \caption{Bitmap images of the umbral core regions around the $S$-light bridge (Identified with $I \leq 0.075I_\text{QS}$ and $|B| > 2$kG ), marked 1, 2 and 3. The umbral contour ($I<0.09I_\text{QS}$) is marked in red. The footpoints (2 \& 3) move in the anti-clockwise direction. }
            \label{fig:umbral-core-frames}
        \end{figure*}
        
        Previous studies define a mature umbra as a region with dark nucleus with hot field-free plasma between them, appearing as UDs or  LBs \citep{Garcia1987}. Therefore, to show the coronal loop originating within the umbra, as stated in \S.~\ref{section:observations}, we searched within the umbra for pixels with intensities, $I < 0.09I_\text{QS}$, to identify the umbral core regions around the LBs. Since the magnetic field of a sunspot umbra ranges from $1800-3700$G \citep{Livingston2002}, we choose a threshold of magnetic field intensity at $2000$ G and an intensity threshold of $0.075 I_\text{QS}$ to identify `core umbra' regions within the sunspot umbra. These values were chosen such that the core umbra that match these conditions do not disrupt or intersect with the light bridges throughout their evolution. \\
        
\begin{figure}
            \centering
            \includegraphics[width=1\linewidth]{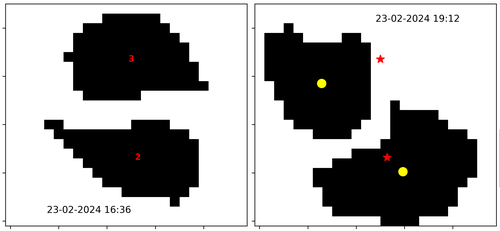}
            \caption{A magnified view of the umbral cores 2 and 3 from Fig.~\ref{fig:umbral-core-frames}. The numbers (red) in the right frame are positioned at the weighted centroid positions. The same positions (initial position of centroids at 16:36) are marked `$\star$' in red, and the final positions of the weighted centroids (at 19:12) are marked in  yellow dots in the right frame. The displacement in the centroids is calculated and velocities deduced for each. Each tick marks 5 pix.}
            \label{fig:umbral-core-vel}
        \end{figure}

        In the case of the $S$-light bridge, we identify 3 regions within the umbra as core umbral regions (see Fig.~\ref{fig:umbral-core-frames}).  
        Of these three regions, regions 2 and 3 form on either side of the LB. When the apparent transverse motion of the LB was observed, these two regions can also be observed moving in a counter-clockwise direction. The weighted centroids of these two regions were tracked over time to study their motion (differential rotation component is removed through de-rotation). We found that between 16:36 to 19:12 UT, the centroids of regions 2 and 3 moved by $2.20$ and $6.60$ pixels, respectively (see. Fig.~\ref{fig:umbral-core-vel}). These shifts result in displacements of $800$km and $2403$km, respectively. These displacements over 4 hours gave us their apparent transverse motion speeds at $0.085$ $kms^{-1}$ and $0.256$ $kms^{-1}$. We correlate these movements with the dynamics of footpoints of the coronal loop, as discussed in \S~\ref{section:coronalobs}. \\
        
         \begin{figure}
            \centering
            \includegraphics[width=1\linewidth]{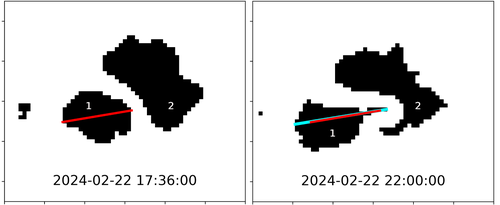}
            \caption{ The umbral core regions of AR 13590 on 22 Feb, around the $\Omega$-light bridge at 17:36 and 22:00. A line is drawn along extended length of the umbral core region to measure the change in width (red line). The extension is shown in cyan. Each tick marks 10 pix.}
            \label{fig:13590-22-len}
        \end{figure}

        The umbral core regions around the $\Omega$-light bridge were identified as two large regions on either side of the LB (see. Fig.~\ref{fig:13590-22-len}). The one on the left side of the LB extends in length along the direction perpendicular to the LB length. We find that between the formation and disappearance of the $\Omega$-light bridge, the two umbral cores did not move significantly. Therefore, we measured the change in the width of the left umbral core along the direction of the `push', from 17:36 to 22:00 UT. The width of the umbral core, in that direction, increases from $17.44$ pixels to $22.99$ pixels. These values translate from $6353$km to $8377$km, respectively. Therefore, the umbral core expands by $2$Mm, perpendicular to the LB. \\

\section{Discussion \& Conclusions}
\label{section:discussion}

    Light bridges are believed to be protrusions of field-free plasma into the magnetized umbra \citep{Parker1979, Rimmele2008, Leka1997, Toriumi2015}. Within the umbra, the field lines are tightly packed, and an LB cannot push the magnetic flux tubes within the umbra \citep{Parker1979}. Consequently, we surmise that the observed transverse motion within the umbra seen in these light bridges can be explained if the motion were driven by magnetic flux tubes in the umbra. For instance, kink MHD waves ($m=1$) are known to displace the flux tube axis \citep{Gossens2009}. The excitation or de-excitation of these kink modes could result in the adjacent flux tubes exerting magnetic pressure on nearby loops. Nearby flares are also known to cause transverse oscillations in coronal loops \citep{Aschwanden1999}. \\
    
    \subsection{Umbral filaments and Sunspot scars}

    In the early stages of their formation, the light bridges observed show properties similar to those of an umbral filament-type LB. The field strength across its width drops, and the field inclination with respect to the normal to the solar surface increases (see Fig.~\ref{fig:inclination-field-plot}). The Dopplergrams also reveal CEFs at the umbra-penumbra boundary, where the LB protrudes into the umbra (see Fig.~\ref{fig:dopplergrams} \& ~\ref{fig:dopplergrams_22}). These observed properties of the light bridges match the description of `umbral filaments' studied by \citet{Kleint2013}, in particular, the penumbral intensity protrusion, lack of granulation, and the magnetic field properties. The umbral filaments were also observed at chromospheric heights, so does the $S$-light bridge \textit{arm} (see Fig.~\ref{fig:chromosphere-plots}). However, the \textit{tail} of the $S$-light bridge and the entire $\Omega$-light bridge were not visible at chromospheric wavelengths (AIA 1600 \r{A} and AIA 1700 \r{A}) in our observations. \\

     The $S$-light bridge \textit{arm} before the formation of the \textit{tail} shows a pattern (left and right columns in Fig.~\ref{fig:inclination-field-plot}) similar to that of sunspot scars \citep{Xing2024}. These authors also report that the inner curve of the sunspot scar is where the flux rope can be found, which holds good in case of both LBs we observed here. The two curves of the $S$-light bridge, both have coronal loop footpoints (see Fig.~\ref{fig:coronal-loop-large}) and the $\Omega$-light bridge was also curved towards the coronal loop base (see Fig.~\ref{fig:coronal-footpoint-22}). \\

    The umbral filaments, that were identified as the footpoints of coronal loops, are said \citep{Kleint2013} to influence the structure of these coronal loops associated with flares. Flare ribbons have also been observed in the vicinity of the sunspot scars \citep{Xing2024}. AR 13590 was one of the most flaring active regions of the year (see Table~\ref{table:flare-table}). These similarities between umbral filaments, sunspot scars and our light bridges and their association with flares, match well with previously reported results. We believe that further studies are required to clearly distinguish between various forms of LBs and their association with coronal activities. \\

    \begin{table}
        \centering
        \begin{tabular}{lc}
        \hline
        \multicolumn{2}{c}{22 Feb} \\
        Flares (class) \& Event & Time \\
        \hline
        
        M1.5    & 00:14 \\
        C3.9    & 03:47 \\
        2 LBs form & 04:12 \\
        X1.7    & 06:32 \\
        C3.4    & 11:39 \\
        C3.3    & 15:35 \\
        LBs combine $\Omega$ LB forms & 16:00 \\
        C9.3    & 16:29 \\
        C4.6    & 17:16 \\
        C4.5    & 17:57 \\
        $\Omega$-LB protrusion begins & 18:24 \\
        C4.7    & 18:25 \\
        M4.8    & 20:46 \\
        X6.37   & 22:34 \\
        $\Omega$-LB max protrusion & 22:36 \\
        $\Omega$-LB disappears & 22:48 \\
        
        \hline \\
        \multicolumn{2}{c}{23 Feb} \\
        Flares (class) \& Event & Time \\
        \hline
        
        C6.4    & 00:54 \\
        $S$-LB \textit{arm} protrudes & 01:12 \\
        C5.6    & 01:27 \\
        C3      & 05:44 \\
        C8.6    & 06:33 \\
        C3.7    & 09:54 \\
        C3.7    & 10:15 \\
        $S$-LB touches opp. boundary & 12:48 \\
        M1      & 13:28 \\
        $S$-LB transverse motion begins & 14:48 \\
        M1.5    & 15:53 \\
        M2.6    & 17:47 \\
        $S$-LB max size & 19:00 \\
        $S$-LB \textit{tail} disappears & 20:00 \\
        C5.4    & 20:39 \\
        C6      & 21:15 \\
        C4.9    & 23:13 \\
        
        \hline
        \end{tabular}
        \caption{Timestamps of flares and LB events for AR 13590 on 22 and 23 Feb 2024.}
        \label{table:flare-table}
        \end{table}

    \subsection{Apparent Transverse Motion}

\begin{figure}
        \centering
        \includegraphics[width=1\linewidth]{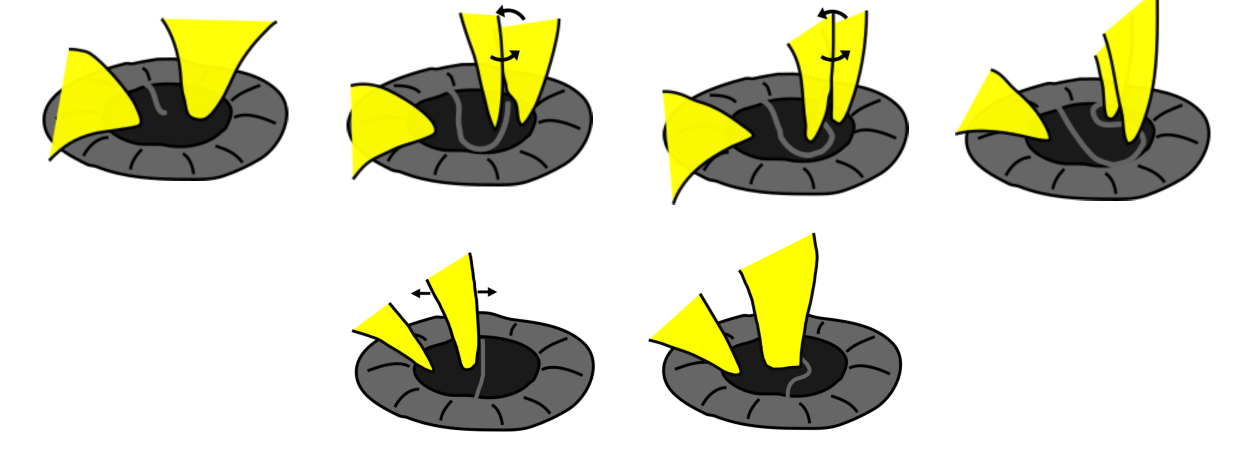}
        \caption{A cartoon of the coronal loop footpoints (yellow) within AR 13590; \textit{top}: around the $S$-light bridge, the umbral cracks (marked by 'gray' line within 'black' umbra) forms around the two footpoints, passing in between them; \textit{bottom}: around the $\Omega$-light bridge, the coronal loop footpoint expands laterally pushing the umbral cracks.}
        \label{fig:cartoon}
    \end{figure}

        The two light bridges observed in AR 13590 exhibit apparent transverse motions. To understand  these motions, we studied the dynamics of the umbral core flux tubes surrounding the LBs (\S \ref{section:umbralcore}). 
        These dynamics reveal that in the case of the $\Omega$-light bridge, the change in the morphology of one of the umbral cores drives the light bridge motion. In the case of the $S$-light bridge, the umbra core regions (regions 2 \& 3 in \ref{fig:umbral-core-frames}) move in counter-clockwise directions, pushing the curved structures of the LB, which may be causing the light bridge's `$S$' shape to broaden, thus masquerading as an apparent transverse motion of the LB. We also noticed from Fig.~\ref{fig:stplot} that this motion of the umbral core regions, which are the footpoints of the coronal loop, causes the transverse motion of the coronal loop. This effect is illustrated in the cartoon depicted in Fig.~\ref{fig:cartoon}. \\
        
        In the case of the $\Omega$-light bridge, the coronal loop near the LB broadens as shown in Fig. \ref{fig:stplot22}, which corresponds to the extension of the umbral core regions in that direction. This `push' by the umbral core region deforms the LB into the shape we identify it with, as illustrated in the cartoon in Fig.~\ref{fig:cartoon}.  This observation suggests that the LB dynamics is coupled with the coronal loops around them.\\
        
        Table \ref{table:flare-table} shows that both LBs evolved around the occurrences of flares. \citet{Thaler2023}, in their simulations, have shown that the footpoints of magnetic loops move sideways during reconnection events. \\
    
        We observed in Fig.~\ref{fig:coronal-loop-small} that the coronal loop footpoints separate and the LB passes between them. Coronal loop bifurcation due to LB formation within an umbra has been previously observed \citet{Miao2021}. The light bridges observed in this study show good temporal correlation between their evolution and the motion of the coronal loop footpoints, but no signatures in the chromosphere. Therefore, we believe that these LBs are photospheric or sub-photospheric in nature. These light bridges can be manifestations of unmagnetized plasma within tiny `umbral cracks' - which are the regions between the core umbra. This explains why the magnetic field intensity and its components have values similar to those of the surrounding umbra in the same region, in the absence of the tail (See Fig.~\ref{fig:inclination-field-plot}).  This could be a consequence of the limited resolution of HMI, which is unable to resolve the magnetic fields intensities along the umbral crack. However, the subsurface nature of the umbral crack could also lead to the convergence of field lines, forming a cusp-like structure above the crack. This would explain the decrease in the magnetic field inclination across the tail. \\

        Due to the multiple flares and reconnection events that occur in this region (see Table~\ref{table:flare-table}), the flux tubes within the core umbra can move around \citep{Thaler2023}. As they move around these, the shapes of these cracks change, and they appear as apparent transverse motions of such faint LBs, and these cracks can also merge and squeeze out the plasma, leading to the disappearance of the light bridges. The existence of these umbral cracks at deeper layers can explain why these light bridges are very faint and narrow, and therefore lower the inclination of the magnetic field. This can also explain why the light bridges are not visible at chromosphere heights. This also hints at these umbral cracks being sub-surface structures. The presence of CEFs (or the lack of Evershed flow) exactly at the point where the LB protrudes from the umbra-penumbra boundary suggests that the CEFs observed are due to flow of plasma from the penumbra into these umbral cracks. We also observed that, as the $\Omega$-LB disintegrated, the upper half remains intact although not static. The \textit{arm} of the $S$-LB that showed conventional light bridge values could be due to the large amount of plasma that initially \textit{flows} into the umbral crack, which then settles as the crack grows into the $S$ shape (see frames 1b, 2b and 3b in Fig.~\ref{fig:cont_enhanced}). 
        Presence of CEFs at the umbral boundary have been indicators of fragmentation of the umbra, beginning from those points on the boundary \citep{Louis2014}.Umbral cracks may grow and lead to sunspot fragmentation. However, in the two light bridges studied here, the cracks failed to develop sufficiently, likely due to the merging of the umbral core regions. Currently, we cannot  determine what causes these cracks. Further investigation of more sunspots and such similar LBs is required to determine the cause for the appearance and disappearance of such umbral cracks that manifest as these faint LBs. \\

    \subsection{Breaking, Disappearance and Coronal Coupling}

        Our analysis of the umbral core regions leads us to a point in the LB evolution where they both disappear as the umbral core regions merge into one. However, in the case of the $S$-light bridge small coronal loops  appear exactly at the time the LB \textit{arm} and the \textit{tail} break apart. A very similar loop also becomes visible when the \textit{tail} disappears abruptly (see Fig.~\ref{fig:coronal-loop-cea}). We currently cannot determine how the appearance of these loops correlates with the LB events. However, such LB events are indeed related to the formation of coronal bright loops, further establishing the coupling between coronal loops and LB dynamics. Light-bridge events have been reported to influence the formation of coronal loops \citep{Feng2020}. Recent advances in direct measurement of magnetic fields at chromosphere (using NLTE Inversion) has enabled us to estimate magnetic field vectors from deep photosphere upto upper chromosphere. Such high resolution multi-height spectrpolarimetric observations along with co-spatial observations of coronal loop footpoints, would provide a better understanding of faint light bridges and umbral magnetic flux tubes. \\

\begin{acknowledgments}
The authors thank the anonymous referee for the useful comments and suggestions. The authors acknowledge the open data provided by the Join Science Operations Center at JSOC, courtesy of NASA/SDO and the AIA and HMI science teams. A.B. acknowledges Dr. T.M.A Pai Fellowship offered for the Ph.D. program from Manipal Academy of Higher Education (MAHE), Manipal. A.B. also acknowledges P.S. Athiray, UAH, USA, and Shin Toriumi, JAXA, JP, for their insights and suggestions. The Manipal Centre for Natural Sciences, MAHE is acknowledged for its facilities and support.
\end{acknowledgments}

\facility{SDO(HMI and AIA)}

\software{astropy \citep{astropy},  
          scipy \citep{scipy},
          numpy \citep{numpy},
          matplotlib \citep{Matplotlib}
          }

\bibliography{references}{}
\bibliographystyle{aasjournalv7}

\end{document}